\documentclass[prl,aps,twocolumn,tightenlines,showpacs,superscriptaddress]{revtex4}
\usepackage{graphics}
\usepackage{color}
\usepackage{epsf}

\usepackage{relsize}

\def\cp{$CP$\/}
\def\mbc{$M^{}_{\rm bc}$}
\def\mmbc{M^{}_{\rm bc}}
\def\deltaE{$\Delta E$}

\def\phitwo{$\phi^{}_2$}

\def\meve{~MeV}

\def\gevp{~GeV/$c$\/}

\def\ra{\!\rightarrow\!}
\def\bbar{\overline{B}{}^{\,0}}

\def\brhorho{$B^0\ra\rho^+\rho^-$}

\def\arhorho{${\cal A}$}
\def\srhorho{${\cal S}$}

\def\babar{\mbox{\slshape B\kern-0.1em{\smaller A}\kern-0.1em
    B\kern-0.1em{\smaller A\kern-0.2em R}}}

\begin{document}

\hspace{14cm} \hbox{\bf BELLE Preprint 2007-21}

\hspace{14cm} \hbox{\bf KEK Preprint 2007-9}

\hspace{14cm} \hbox{\bf BC-0678}

\hspace{14cm} \hbox{\bf UCHEP-07-02}

\vspace{1.2cm}

\title{ 
{\boldmath Improved measurement of $CP$-violating parameters in
\brhorho decays}}

\affiliation{Budker Institute of Nuclear Physics, Novosibirsk}
\affiliation{University of Cincinnati, Cincinnati, Ohio 45221}
\affiliation{Department of Physics, Fu Jen Catholic University, Taipei}
\affiliation{The Graduate University for Advanced Studies, Hayama}
\affiliation{Hanyang University, Seoul}
\affiliation{University of Hawaii, Honolulu, Hawaii 96822}
\affiliation{High Energy Accelerator Research Organization (KEK), Tsukuba}
\affiliation{Institute of High Energy Physics, Chinese Academy of Sciences, Beijing}
\affiliation{Institute of High Energy Physics, Vienna}
\affiliation{Institute of High Energy Physics, Protvino}
\affiliation{Institute for Theoretical and Experimental Physics, Moscow}
\affiliation{J. Stefan Institute, Ljubljana}
\affiliation{Kanagawa University, Yokohama}
\affiliation{Korea University, Seoul}
\affiliation{Kyungpook National University, Taegu}
\affiliation{Swiss Federal Institute of Technology of Lausanne, EPFL, Lausanne}
\affiliation{University of Ljubljana, Ljubljana}
\affiliation{University of Maribor, Maribor}
\affiliation{University of Melbourne, School of Physics, Victoria 3010}
\affiliation{Nagoya University, Nagoya}
\affiliation{Nara Women's University, Nara}
\affiliation{National Central University, Chung-li}
\affiliation{National United University, Miao Li}
\affiliation{Department of Physics, National Taiwan University, Taipei}
\affiliation{H. Niewodniczanski Institute of Nuclear Physics, Krakow}
\affiliation{Nippon Dental University, Niigata}
\affiliation{Niigata University, Niigata}
\affiliation{Osaka City University, Osaka}
\affiliation{Osaka University, Osaka}
\affiliation{Panjab University, Chandigarh}
\affiliation{Peking University, Beijing}
\affiliation{Princeton University, Princeton, New Jersey 08544}
\affiliation{RIKEN BNL Research Center, Upton, New York 11973}
\affiliation{University of Science and Technology of China, Hefei}
\affiliation{Seoul National University, Seoul}
\affiliation{Sungkyunkwan University, Suwon}
\affiliation{University of Sydney, Sydney, New South Wales}
\affiliation{Tata Institute of Fundamental Research, Mumbai}
\affiliation{Toho University, Funabashi}
\affiliation{Tohoku Gakuin University, Tagajo}
\affiliation{Tohoku University, Sendai}
\affiliation{Department of Physics, University of Tokyo, Tokyo}
\affiliation{Tokyo Institute of Technology, Tokyo}
\affiliation{Tokyo Metropolitan University, Tokyo}
\affiliation{Tokyo University of Agriculture and Technology, Tokyo}
\affiliation{Virginia Polytechnic Institute and State University, Blacksburg, Virginia 24061}
\affiliation{Yonsei University, Seoul}
  \author{A.~Somov}\affiliation{University of Cincinnati, Cincinnati, Ohio 45221} 
  \author{A.~J.~Schwartz}\affiliation{University of Cincinnati, Cincinnati, Ohio 45221} 
  \author{I.~Adachi}\affiliation{High Energy Accelerator Research Organization (KEK), Tsukuba} 
  \author{H.~Aihara}\affiliation{Department of Physics, University of Tokyo, Tokyo} 
  \author{D.~Anipko}\affiliation{Budker Institute of Nuclear Physics, Novosibirsk} 
  \author{V.~Aulchenko}\affiliation{Budker Institute of Nuclear Physics, Novosibirsk} 
  \author{T.~Aushev}\affiliation{Swiss Federal Institute of Technology of Lausanne, EPFL, Lausanne}\affiliation{Institute for Theoretical and Experimental Physics, Moscow} 
  \author{A.~M.~Bakich}\affiliation{University of Sydney, Sydney, New South Wales} 
  \author{I.~Bedny}\affiliation{Budker Institute of Nuclear Physics, Novosibirsk} 
  \author{U.~Bitenc}\affiliation{J. Stefan Institute, Ljubljana} 
% ...
\collaboration{The Belle Collaboration}

\noaffiliation

\begin{abstract}
We present a measurement of the \cp-violating asymmetry in \brhorho\ decays
using 535 million $B\overline{B}$ pairs collected with the Belle detector
 at the KEKB $e^+ e^-$ collider. We measure \cp-violating coefficients 
${\cal A} = 0.16\,\pm 0.21\,({\rm stat})\,\pm 0.08\,({\rm syst})$ and
 ${\cal S} = 0.19\,\pm 0.30\,({\rm stat})\,\pm 0.08\,({\rm syst})$.
These values are used to determine the unitarity triangle angle 
$\phi_2$ using an isospin analysis; the solution consistent with the Standard Model 
lies in the range $54^\circ\!<\!\phi^{}_2\!<\!113^\circ$ at the 90\% confidence
level.

\end{abstract}


\maketitle

$CP$ violation in the Standard Model is attributed to the presence of 
an irreducible complex phase in the Cabibbo-Kobayashi-Maskawa~\cite{ckm} 
(CKM) quark-mixing matrix. The unitarity of the CKM matrix leads to six
triangles in the complex plane. One such triangle is given by
the following relation among the  matrix elements: 
$V_{ud}V^*_{ub} + V_{cd}V^*_{cb} + V_{td}V^*_{tb} = 0$.
The phase angle  $\phi_2$, defined as ${\rm arg}[-(V_{td}V_{tb}^*)/(V_{ud}V_{ub}^*)]$, 
can be determined by measuring a time-dependent $CP$ asymmetry in 
$b\rightarrow u\overline{u}d$ decays such as $B^0\ra\pi^+\pi^-,\,(\rho\pi)^0$, 
and $\rho^+\rho^-$~\cite{chargeconjugate}. The time-dependent rate for 
$B \rightarrow \rho^+\rho^-$  decays tagged with $B^0$($q=+1$) and $\bbar$ ($q = -1$) 
mesons is given by
\begin{eqnarray}
\mathcal{P}_{\rho\rho} (\Delta t) =  \frac{e^{-|\Delta t|/\tau_{B^0}}}{4\tau_{B^0}}
\{ 1  + q[{\cal A}_{}\cos(\Delta m \Delta t) \\
 + {\cal S}_{}\sin(\Delta m \Delta t)]\},\nonumber
\label{eqn:rate}
\end{eqnarray}
where $\tau_{B^0}$ is the $B^0$ lifetime, $\Delta m$ is the mass difference between 
the two $B^0$ mass eigenstates, $\Delta t$ is the proper-time difference between
the two $B$ decays in the event, and ${\cal A_{}}$ and ${\cal S_{}}$ are $CP$ asymmetry 
coefficients. If the decay amplitude is a pure $CP$-even state  and is dominated by a tree diagram, 
 ${\cal S}_{} = {\rm sin}(2\phi_2)$ and ${\cal A}_{} = 0$. The presence 
of an amplitude with a different weak phase (such as from a loop diagram) gives rise 
to direct $CP$ violation and shifts ${\cal S}_{}$ from ${\rm sin}(2\phi_2)$. 
However, the size of a loop amplitude is constrained to be small by the small branching fraction 
of $B^0\ra\rho^0\rho^0$~\cite{babar_rho0rho0}.

The $CP$-violating parameters receive contributions from a longitudinally polarized 
state ($CP$-even) and two transversely  polarized states (an admixture of $CP$-even and 
$CP$-odd states). Recent measurements of the polarization fraction by Belle~\cite{belle_rhorho} and 
BaBar~\cite{babar_alpha} show that the longitudinal polarization fraction is near unity 
($f_L=0.968\pm 0.023$~\cite{hfag}).

Here we present an improved measurement of the $CP$-violating coefficients \arhorho\ and 
\srhorho\ using $492\,{\rm fb^{-1}}$ of data containing 535 million $B\overline{B}$ pairs. 
This data sample is about a factor of two larger than 
that used in  our earlier publication~\cite{belle_rhorho}. In addition we have modified the event 
selection by reducing the threshold on a continuum suppression variable; this increases our 
reconstruction efficiency by about 70$\%$. We subsequently introduce a probability density 
function (PDF) for the continuum suppression variable into the likelihood function; this provides 
additional discrimination between signal and backgrounds.
The expected improvement in the statistical error of ${\cal A}_{}$ and  
${\cal S}_{}$ due to the new event selection is about $12\%$.

The  $B\overline{B}$ pairs were collected with the Belle detector~\cite{belle_detector} at 
the KEKB~\cite{kekb} $e^+e^-$ asymmetric-energy (3.5 GeV on 8.0 GeV) collider with a 
center-of-mass (CM) energy at the $\Upsilon(4S)$ resonance. 
The $\Upsilon(4S)$ is produced with a Lorentz boost of $\beta \gamma = 0.425$ nearly along 
the $z$ axis, which is oriented antiparallel to the positron beam. Since the $B^0$ and $\bbar$ 
mesons are produced approximately at rest in the $\Upsilon(4S)$ CM system, the decay time 
difference $\Delta t$ is related to the distance between the decay vertices of the two $B$ 
mesons as $\Delta t \simeq \Delta z / \beta \gamma c$, where $c$ is the speed of light.

The Belle detector~\cite{belle_detector} is a large-solid-angle spectrometer.
It includes a silicon vertex detector (SVD), a 50-layer central drift chamber (CDC),
an array of aerogel threshold  Cherenkov counters (ACC), time-of-flight scintillation
counters (TOF), and an electromagnetic calorimeter (ECL)  comprised of CsI(Tl) crystals located
inside a superconducting solenoid coil that provides a 1.5 T magnetic field.

We reconstruct \brhorho decays by combining two oppositely charged 
pion tracks with two neutral pions. Each charged track is required to have a 
transverse momentum $p^{}_T\!>\!0.10$\gevp\ in the laboratory frame  and originate 
within $dr\!<\!0.2$~cm in the radial direction and within $|dz|\!<\!4.0$~cm along the $z$-axis 
from the interaction point, which is determined run-by-run. 
A track is identified as a pion using information from the CDC, ACC and TOF systems. 
Tracks matched with clusters in the ECL that are consistent with an
electron hypothesis are rejected.

The $\pi^0$ candidates are reconstructed from $\gamma\gamma$ pairs with an invariant mass 
in the range $117.8~{\rm MeV}/c^2\!<\!M^{}_{\gamma\gamma}\!<\!150.2~{\rm MeV}/c^2$ 
(about $\pm 3\sigma$ in $m_{\pi^0}$ resolution). Photons are required to have energy 
$E^{}_\gamma\!>\!50$\meve\ in the ECL barrel region ($32^\circ\!<\!\theta\!<\!129^\circ$) 
and $E^{}_\gamma\!>\!90$\meve\ in the endcap regions ($17^\circ\!<\!\theta\!<\!32^\circ$ 
and $129^\circ\!<\!\theta\!<\!150^\circ$), where $\theta$ denotes the polar angle with 
respect to the $z$ axis.

\begin{figure}[t]
\mbox{\epsfxsize=3.4in \epsfbox{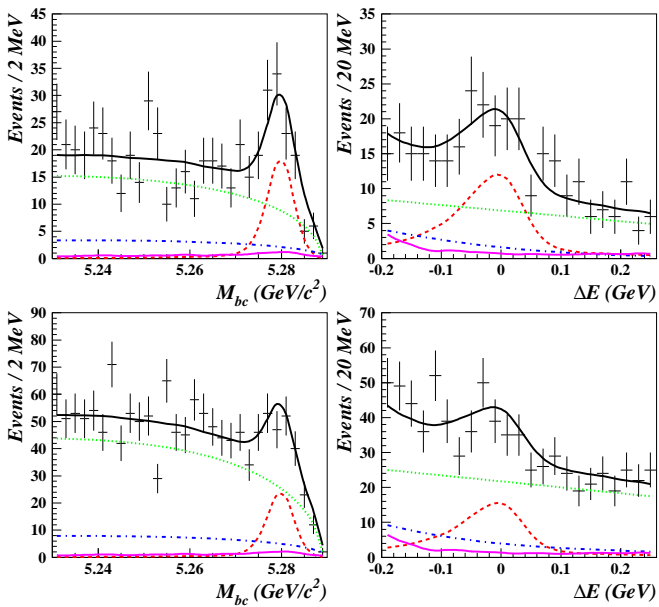}}
\caption{ Left: projections in \mbc\ for events satisfying 
$-0.10~{\rm GeV}\!<\!\Delta E\!<\!0.06~{\rm GeV}$. Right: projections in \deltaE\ for 
events with $5.27~{\rm GeV}/c^2\!<\!M_{\rm bc}\!<\!5.29~{\rm GeV}/c^2$. 
The top plots correspond to good quality tags ($0.75\!<\!r\!<\!1.0$), and the 
bottom plots correspond to lower  quality tags ($r\!<\!0.75$). 
The curves show fit projections: dashed is $\rho^+\rho^-\!+\rho\pi\pi$,
dotted is $q\bar{q}$, dot-dashed is $b\ra c$, small solid is $b\ra u$, and 
large solid is the total. For these plots the ${\cal R}$ requirement has been 
tightened to increase the ratio of signal to background.}
\label{fig:one}
\end{figure}
\begin{figure}[t]
\mbox{\epsfxsize=3.4in \epsfbox{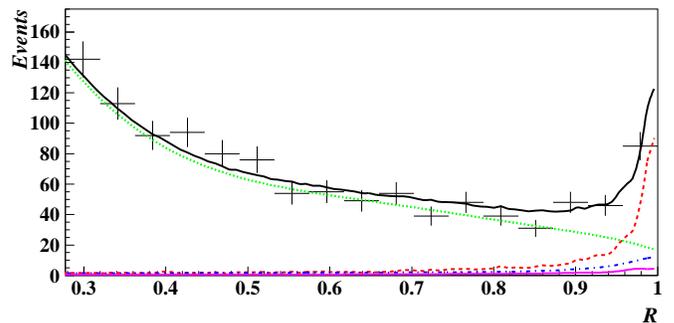}}
\caption{{$\cal R$} distribution for high-purity tagged events satisfying 
$5.27~{\rm GeV}/c^2\!<\!M_{\rm bc}\!<\! 5.29~{\rm GeV}/c^2$ and 
$-0.10~{\rm GeV}\!<\!\Delta E\!<\!0.06~{\rm GeV}$. The curves show fit projections: 
dashed is $\rho^+\rho^-\!+\rho\pi\pi$, dotted is $q\bar{q}$, dot-dashed is $b\ra c$, 
small solid is $b\ra u$, and large solid is the total. }
\label{fig:two}
\end{figure}
To reconstruct $\rho^\pm$ mesons,  we combine $\pi^\pm$ candidates with $\pi^0$ 
candidates. The $\pi^\pm\pi^0$  combination must have an invariant mass in the 
range $0.62~{\rm GeV}/c^2\!<\! M_{\pi^\pm\pi^0}\!<\!0.92~{\rm GeV}/c^2$. 
To reduce combinatorial background, we reject $\pi^0$'s with $p<0.35$\gevp\, 
in the CM frame. We also require  $-0.80\!<\!\cos\theta^{}_{\!\pm}\!<\!0.98$, 
where $\theta^{}_{\!\pm}$ is the angle between the direction of the $\pi^0$ from 
the $\rho^\pm$ and the negative of the $B^0$ momentum in the $\rho^\pm$ rest frame.

\brhorho\ decays are identified using the beam-energy-constrained mass 
$\mmbc\!\equiv\!\sqrt{E^2_{\rm beam}-p^2_B}$ and energy difference 
$\Delta E\!\equiv\!E^{}_B-E^{}_{\rm beam}$, where $E^{}_{\rm beam}$ is the  
beam energy, and $E^{}_B$ and $p_B$ are the energy and momentum of the 
reconstructed $B$  candidate, all evaluated in the CM frame.

The flavor of the $B$ meson accompanying the \brhorho\ candidate is identified 
via its decay products: charged leptons, kaons, and $\Lambda$'s. A tagging 
algorithm~\cite{tag_algorithm} yields the flavor of the tagged meson, $q$, and 
a quality factor, $r$. The parameter $r$ ranges from 0 for no flavor 
discrimination to 1 for unambiguous flavor assignment. We divide the  data sample 
into six $r$ intervals (denoted  $\ell$=1,2,...,6). The wrong tag 
fractions $\omega_\ell$ for these intervals and the differences $\Delta \omega_\ell$ 
in these fractions between $B^0$ and $\bbar$ decays are determined from 
data~\cite{tag_algorithm}.

The dominant background originates from  $e^+e^-\!\ra q\bar{q}\ (q=u,d,s,c)$
continuum events. To separate $q\bar{q}$ jet-like events from more spherical 
$B\overline{B}$ events, we use event-shape variables, specifically, 16 modified 
Fox-Wolfram moments~\cite{fox_wolfram} combined into a Fisher discriminant~\cite{KSFW}. 
We form signal and background likelihood functions ${\cal L}^{}_s$ and 
${\cal L}^{}_{BG}$ by multiplying the PDF for the Fisher discriminant by a PDF for 
$\cos\theta^{}_B$, where $\theta^{}_B$ is the polar angle in the CM frame between the 
$B$ direction and the beam axis. The PDFs for signal and $q\bar{q}$ are obtained from 
Monte Carlo (MC) simulations and the data sideband 
$5.23~{\rm GeV}/c^2\!<\! M_{\rm bc}\!<\!5.26~{\rm GeV}/c^2$, respectively. 
We calculate the ratio ${\cal R} ={\cal L}^{}_s/({\cal L}^{}_s + {\cal L}^{}_{BG})$ and 
make a loose requirement ${\cal R}\!>\!0.15$. 

The decay vertices of a $\rho^+\rho^-$ candidate and the tag-side $B$ meson
are reconstructed using charged tracks that have a sufficient number of SVD 
hits and an interaction point constraint. The vertex reconstruction algorithm is
described in Ref.~\cite{resolution}.

The analysis is organized in two main steps. We first determine the
yields of signal and background components from a fit to the three-dimensional 
$(\mmbc,\Delta E, {\cal R})$ distribution. Here, $B^0$ candidates are required 
to satisfy $5.23~{\rm GeV}/c^2\!<\!\mmbc\!<\!5.29~{\rm GeV}/c^2$, 
$-0.2~{\rm GeV}\!<\!\Delta E\!<\!0.26~{\rm GeV}$, and ${\cal R}\!>\!0.15$.  We 
subsequently perform a fit to the $\Delta t$ distribution to determine the $CP$ parameters 
\arhorho\ and \srhorho. The signal region used for the $\Delta t$ fit is 
$5.27~{\rm GeV}/c^2\!<\!\mmbc\!<\!5.29~{\rm GeV}/c^2$, 
$-0.12~{\rm GeV}\!<\!\Delta E\!<\!0.08~{\rm GeV}$, and ${\cal R}\!>\!0.15$.

About $12.6\,\%$ of events contain multiple \brhorho\ candidates, most of which 
arise from fake $\pi^0$'s combining with good tracks. We select the best candidate 
based on the $\pi^0$ masses, i.e., minimizing $\sum_{\pi^0_{1,2}}(m_{\gamma\gamma} - m_{\pi^0})^2$. 
For the small fraction ($3\%$) of multiple-candidate events that arise due to extraneous $\pi^\pm$ 
tracks combining with a single $\pi^0$, we select one randomly. Signal decays that have at 
least one $\pi$ meson incorrectly identified are referred to as ``self-cross-feed'' (SCF) events.

The likelihood function used to determine the event yields is given by
\begin{eqnarray}
{\mathcal L} = {\rm exp}\left(-\sum_j N_j\right)\prod_{i=1}^{N_{\rm evt}}\left[\sum_j N_j {\mathcal P_j(M^{i}_{\rm bc},{\Delta E}^i, 
{\cal R}^i)}\right],
\end{eqnarray}
where $j$ indicates one of the following event categories: signal and $\rho\pi\pi$ 
non-resonant decays, SCF events, continuum background ($q\bar q$), $b\rightarrow c$ background, and charmless 
($b\to u$)  background. $N_j$ is the yield of each category, ${\mathcal P}_j(M^{i}_{\rm bc},{\Delta E}^i,{\cal R}^i)$ 
is the PDF for the $i$-th event for category $j$, and $N_{\rm evt}$ is the total number of events in the fit. Except 
for the small contributions of $b\rightarrow u$ background and SCF events, the yields $N_j$ are determined from the fit.
Due to the similar shapes of the $\mmbc$, $\Delta E$, and  ${\cal R}$ distributions for signal and 
$\rho\pi\pi$ events, we cannot distinguish these two components; the fraction of $\rho\pi\pi$ events is 
measured in Ref.~\cite{belle_rhorho} and constitutes $(6.3\pm6.7)\,\%$ of the total 
$N_{\rho^+\rho^- + \rho\pi\pi}$ signal.
The fraction of SCF events is determined from MC 
simulation. The \mbc\ and \deltaE\ shapes for the signal and SCF components are modeled by a 
two-dimensional smoothed histogram obtained from a large MC  sample. 
To take into account a small difference between the MC and data, the \mbc\ - \deltaE\ shapes are 
corrected according to calibration factors determined from a $B^+\rightarrow {\bar{ D}}^0\rho^+$, 
${\bar{ D}}^0\rightarrow K^+\pi^-\pi^0$ control sample. The ${\cal R}$ shapes are modeled by 
one-dimensional histograms, also obtained from MC simulation.

The PDF for $b\ra c$ background is the product of a threshold ARGUS function~\cite{argus} for \mbc, 
a quadratic polynomial for \deltaE, and the sum of a Gaussian and a third-order polynomial for 
${\cal R}$. The shapes of the \deltaE\ and ${\cal R}$ distributions depend on the tag quality 
bin $\ell$. Parameters for all distributions are obtained from the MC.

The \mbc\ and \deltaE\ PDFs for $q\bar{q}$ are modeled by an ARGUS function and a linear function, 
respectively. The \deltaE\ slope depends on ${\cal R}$  and the tag quality bin $\ell$. The shape 
parameters for  \mbc\ and \deltaE\ are floated  in the fit. The ${\cal R}$  PDF for $q\bar{q}$ 
background is taken to be an eighth-order polynomial function; the coefficients depend on the bin 
$\ell$ and are determined from a data sample collected at a CM energy $\sim60~{\rm MeV}$ below 
the $\Upsilon(4S)$.

The $b\ra u$ background is dominated by $B\ra (\rho\,\pi,\,a^{}_1\pi,\,a^{}_1\,\rho)$ decays.
We estimate the $B^\pm\!\rightarrow\!(a_1\pi)^\pm$  branching fractions (which are unmeasured) 
to be $(20 \pm 10) \times 10^{-6}$ using the measured value for 
$B^0\!\rightarrow\!a_1^\pm\pi^\mp$~\cite{ap_pim}. For $B^\pm\!\rightarrow\!(a_1\rho)^\pm$ 
we assume branching fractions of $(30 \pm 15) \times 10^{-6}$, consistent with the present 
upper limit for $B^0\!\rightarrow\!a_1^\pm\rho^\mp$ ($<\!6\times 10^{-5}$~\cite{babar_aprhom}).
The fraction of $b\ra u$ events is very small ($0.37\%$) and thus is 
fixed in the fit according to the prediction of MC simulation. A fit to 176843 events maximizing 
${\mathcal L}$ yields $N_{\rho\rho+\rho\pi\pi} = 576\pm 53$.  Figures~\ref{fig:one} and ~\ref{fig:two} 
show the \mbc, \deltaE, and ${\cal R}$ distributions along with projections of the fit result.

The $CP$-violating parameters \arhorho\ and \srhorho\ are obtained
using an unbinned ML fit to the $\Delta t$ distribution.
The likelihood function for event $i$ is given by
\begin{eqnarray} & {\cal L}^{}_i = & \hspace*{-0.05in}
\sum_{n}\int_{}^{} f_{n}^{}(\vec{x}_{i})\,{\cal P}_{n}^{}(\Delta t^{\prime})^{}\, 
R_{n}(\Delta t^{i},\Delta t^{\prime})\,d\Delta t^\prime,
\end{eqnarray}
where $n$ is one of the six event categories: correctly reconstructed signal, 
SCF events, $\rho\pi\pi$ non-resonant events, $b\rightarrow c$ background, 
$q\overline q$ background, and $b\rightarrow u$ background.
The weights $f_n^{}$ are functions of $\vec{x}\in (\mmbc,\Delta E, {\cal R})$
and are normalized to the event fractions obtained from the $(\mmbc,\Delta E, {\cal R})$ 
fit. The PDFs ${\cal P}_{n}^{}(\Delta t)$ are convolved with the corresponding $\Delta t$ 
resolution functions~$R_n$. Both  $f_n^{}$ and  ${\cal P}_{n}^{}(\Delta t)$  depend on the 
tag quality bin~$\ell$.

The signal PDF is given by Eq.~(1) modified to take into account the effect of 
incorrect flavor assignment: $e^{-|\Delta t|/\tau^{}_{B^0}}/(4\tau^{}_{B^0})\times
\left\{1 - q\Delta\omega^{}_{\ell} + q (1-2\omega^{}_{\ell}) 
\left[\,{\cal A}\cos(\Delta m\,\Delta t) + {\cal S}\sin(\Delta m\,\Delta t)\,\right]\right\}$. 
As the fraction of longitudinal polarization $f_L$ is close to $100\%$, we assume that 
${\cal A} = {\cal A}_L$, ${\cal S} = {\cal S}_L$, and consider the potential contribution 
from a transversely polarized amplitude as a systematic uncertainty. The signal PDF is 
convolved with the same $\Delta t$ resolution function as that used for Belle's 
$\sin2\phi^{}_1$ measurement~\cite{resolution}.

The fraction of SCF events with incorrectly reconstructed vertices is estimated from MC 
simulation to be $(6.5\pm 0.1)\%$ of all signal events. The PDFs ${\cal P}^{}_{\rho\pi\pi}$ 
and ${\cal P}^{}_{\rm SCF}$ are exponential with $\tau\!=\!\tau^{}_B$ and $\tau\!\approx\!0.96$~ps 
(from MC), respectively; these are smeared by a common resolution function. 

The $\Delta t$ PDFs for the backgrounds are modeled as a sum of prompt and exponential 
components: ${\cal P}_k = f_{\delta}^k\delta(\Delta t) + (1-f_{\delta}^k)e^{-|\Delta t|/\tau_k}/2\tau_k$, 
where $k$ represents continuum, $b\rightarrow c$, and $b\rightarrow u$ backgrounds, 
$f_{\delta}^k$ is the fraction of the prompt component, $\delta(\Delta t)$ is the Dirac 
delta function, and $\tau_k$ is an effective lifetime. These PDFs are convolved with a resolution-like function 
$R_k$ parameterized as a sum of two Gaussian functions. Parameters for ${\cal P}_k$ and  $R_k$ are
determined from a data sideband for continuum background and from large MC samples for $b\rightarrow c$ 
and $b\rightarrow u$ backgrounds. To account for small correlations between the shape of the $\Delta t$ 
distribution and ${\cal R}$ for $q\overline q$ background, the parameters are obtained separately for 
low ($0.15\!<\!{\cal R}\!<\!0.75$) and high ($0.75\!<\!{\cal R}\!<\!1.0$) ${\cal R}$ regions.

\begin{figure}[t]
\mbox{\epsfxsize=3.4in \epsfbox{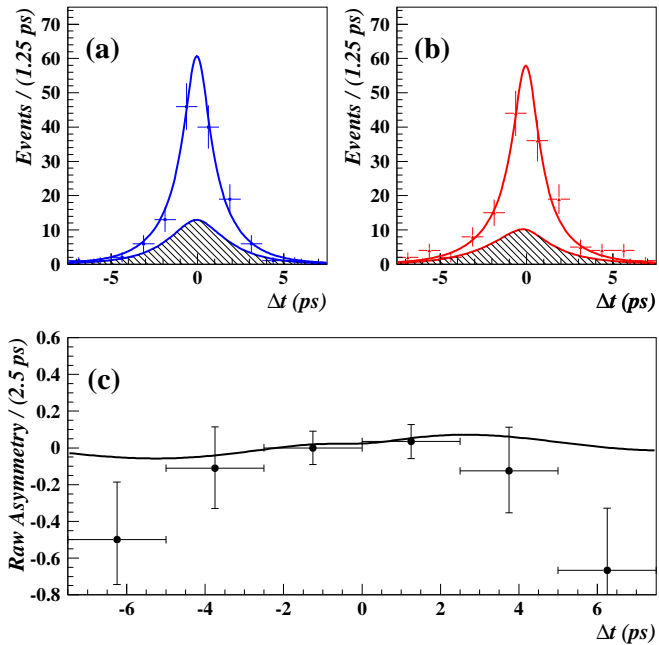}}
\caption{ The $\Delta t$ distribution  and projections of the fit for events 
satisfying $0.5\!<\!r\!<\!1.0$: (a) ~$q\!=\!+1$ tags, (b) ~$q\!=\!-1$ tags. 
The hatched region shows signal events. The raw \cp\ asymmetry is shown in (c). 
For these plots the ${\cal R}$ requirement has been tightened to increase the 
ratio of signal to background.}
\label{fig:three}
\end{figure}

We determine \arhorho\ and \srhorho\ by maximizing $\sum_i \log{\cal L}^{}_i$, where $i$ 
runs over the 18016 events in the $(\mmbc,\Delta E, {\cal R})$ signal region. The results are 
${\cal A}\!=\!0.16\,\pm 0.21$ and ${\cal S}\!=\!0.19\,\pm 0.30$, where the errors are 
statistical. The correlation coefficient is~$-0.10$. These values are consistent with 
no \cp\ violation (${\cal A}={\cal S}=\!0$); the errors are consistent with MC expectations. 
Figure~\ref{fig:three} shows the data and projections of the fit result.

\begin{figure}[t]
\mbox{\epsfxsize=3.4in \epsfbox{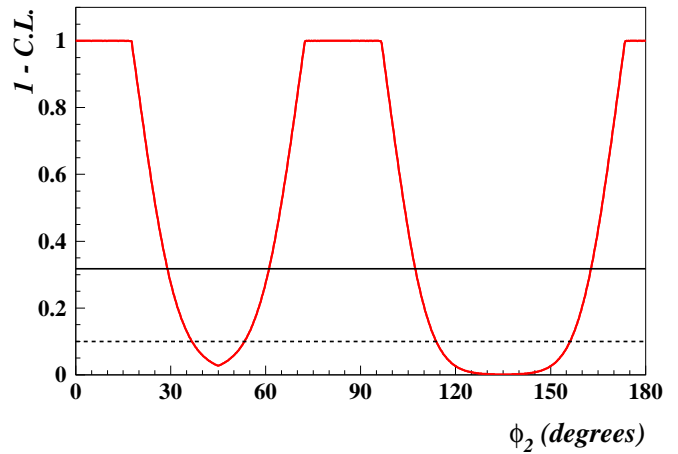}}
\caption{~$1-{\rm C.L.}$ vs. \phitwo. The horizontal lines denote 
${\rm C.L.}\!=\!68.3\%$ (solid) and ${\rm C.L.}\!=\!90\%$ (dashed). } 
\label{fig:four}
\end{figure}

The sources of systematic error are listed in Table~\ref{tab:cpv_systematics}. 
The error for most sources is evaluated by varying the corresponding parameters 
by $\pm 1$ standard deviation ($\sigma$). The effect of a possible asymmetry in 
$b\ra c$ and $q\bar{q}$ is evaluated by adding such an asymmetry to the $b\ra c$ 
and $q\bar{q}$ $\Delta t$ distributions. The uncertainty due to a possible 
asymmetry in $\rho\pi\pi$ non-resonant decays is estimated by varying 
${\cal A}_{\rho\pi\pi}$ and ${\cal S}_{\rho\pi\pi}$ by 0.68, corresponding to 
a $68\%$ confidence interval of a free distribution. We vary the 
branching fractions for $a_1\rho$ and $a_1\pi$ decays and also allow for a $CP$ 
asymmetry of up to $100\%$ in these modes. The error due to transverse polarization 
is obtained by first setting $f^{}_L$ equal to its central value~\cite{hfag} and varying 
${\cal A}^{}_T$, ${\cal S}^{}_T$ from $-1$ to $+1$; then, conservatively 
assuming that the transversely polarized component (with fraction $f_T = 1 - f_L$) is  
pure $CP$-odd for which  ${\cal A}_T = {\cal A}_L$, ${\cal S}_T = -{\cal S}_L$, and varying 
$f_L$ by its error. Summing up in quadrature all systematic uncertainties, we obtain overall 
systematic errors of $\pm\,0.08$ for both ${\cal A}^{}$  and  ${\cal S}^{}$. Thus,

\begin{eqnarray}
{\cal A}^{} & = &
0.16\,\pm 0.21\,({\rm stat})\,\pm 0.08\,({\rm syst}) \\
{\cal S}^{} & = &
0.19\,\pm 0.30\,({\rm stat})\,\pm 0.08\,({\rm syst})\,.
\end{eqnarray}
These values are consistent with, and supersede,  our previous measurement~\cite{belle_rhorho}. 
They are also consistent with results  obtained by BaBar~\cite{babar_alpha}.

\begin{table}[htb]
\caption{Systematic errors for \cp\ coefficients ${\cal A}$ and ${\cal S}$.}
\label{tab:cpv_systematics}
\vskip0.20in
\begin{ruledtabular}
\begin{tabular}{lcccc}
{\bf Type} & \multicolumn{2}{c}{\boldmath $\Delta {\cal A}$ ($\times 10^{-2}$)}   
     & \multicolumn{2}{c}{\boldmath $\Delta {\cal S}$ ($\times 10^{-2}$)} \\
            &  {\boldmath $+\sigma$} & {\boldmath $-\sigma$} & 
{\boldmath $+\sigma$} & {\boldmath $-\sigma$} \\
\hline

Wrong tag fractions                     &   0.5   &   0.5   &   0.8  &  0.8   \\
Parameters $\Delta m$, $\tau_{B^0}$     &   0.2   &   0.3   &   0.6  &  0.7   \\
Resolution function                     &   1.4   &   1.5   &   1.0  &  1.7   \\
Background $\Delta t$ distributions     &   0.5   &   0.5   &   1.0  &  1.1   \\
Component fractions                     &   1.5   &   1.9   &   3.9  &  3.7   \\
$\rho\pi\pi$ non-resonant fractions      &   1.2   &   1.0   &   1.5  &  1.2   \\
SCF fraction, $\Delta t$  PDF           &   0.2   &   0.2   &   0.1  &  0.1   \\
Shape of ${\cal R}$ PDF                 &   0.8   &   0.7   &   1.2  &  1.3   \\
Vertexing                               &   2.1   &   2.1   &   1.0  &  1.3   \\
Possible fitting bias                   &   0.3   &   0.0   &   0.3  &  0.0   \\
Background asymmetry                    &   1.1   &   0.0   &   0.0  &  0.4   \\
$b\rightarrow u$ asymmetry              &   2.4   &   2.9   &   2.4  &  3.2   \\
$\rho\pi\pi$ asymmetry                  &   4.6   &   4.6   &   4.6  &  4.6   \\
Transverse polarization                 &   3.8   &   2.8   &   4.6  &  2.7   \\
Tag-side interference~\cite{Long}       &   3.7   &   3.7   &   0.1  &  0.1   \\
\hline
Total                                   &   8.3   &   8.0   &   8.4  &  7.9   \\
\end{tabular}
\end{ruledtabular}
\end{table}

We constrain $\phi^{}_2$ using an isospin analysis~\cite{gronau_london}, which 
allows one to relate six observables to six underlying parameters: five decay 
amplitudes for $B\ra\rho\rho$  and the angle $\phi_2$. The observables are the 
branching fractions for $B\ra\rho^+\rho^-$, $\rho^+\rho^0$\,\cite{hfag}, and 
$\rho^0\rho^0$\,\cite{babar_rho0rho0}; the $CP$ parameters ${\cal A}$ and ${\cal S}$ 
(our results); and the parameter ${\cal A}_{\rho^0\rho^0}$ for $B\ra\rho^0\rho^0$ decays. 
The last parameter is not yet measured, but nevertheless one can constrain $\phi_2$.
The branching fractions must be multiplied by the corresponding longitudinal polarization 
fractions~\cite{hfag}. We neglect possible contributions from electroweak penguins and  
$I\!=\!1$ amplitudes~\cite{falk} and possible interference between signal and non-resonant 
components. We follow the statistical method of 
Ref.~\cite{charles} and construct a  $\chi^2({\phi_2})$ using the measured values and  
obtain a minimum $\chi^2$ (denoted $\chi^2_{\rm min}$); we then scan \phitwo\ from 0$^\circ$ 
to 180$^\circ$, calculating the difference $\Delta\chi^2\equiv\chi^2(\phi^{}_2)-\chi^2_{\rm min}$. 
We insert $\Delta\chi^2$ into the cumulative distribution function for the $\chi^2$ distribution 
for one degree of freedom to obtain a confidence level (C.L.) for each \phitwo\ value. 
The resulting function $1\!-\!{\rm C.L.}$ (Fig.~\ref{fig:four}) has more than one peak 
due to ambiguities that arise when solving for $\phi^{}_2$. The ``flat-top''  regions in 
Fig.~\ref{fig:four} arise because ${\cal A}_{\rho^0\rho^0}$  is not measured. 
The solution consistent with the Standard Model is $62^\circ\!<\!\phi^{}_2\!<\!106^\circ$ 
at 68\%~C.L. or $54^\circ\!<\!\phi^{}_2\!<\!113^\circ$ at 90\%~C.L. Recently, an alternative 
model-dependent approach to extract $\phi_2$ using flavor $SU$(3) symmetry has been 
proposed~\cite{alpha_su3}. This method could potentially give more stringent constraints 
on $\phi_2$.

In summary, we present an improved measurement of the $CP$-violating coefficients ${\cal A}$ 
and ${\cal S}$ in \brhorho\ decays using $492\,{\rm fb^{-1}}$ of data, which corresponds to 
535 million $B\overline{B}$  pairs. These measurements are used to constrain the angle~$\phi^{}_2$.

We thank the KEKB group for the excellent operation of the
accelerator, the KEK cryogenics group for the efficient
operation of the solenoid, and the KEK computer group and
the National Institute of Informatics for valuable computing
and Super-SINET network support. We acknowledge support from
the Ministry of Education, Culture, Sports, Science, and
Technology of Japan and the Japan Society for the Promotion
of Science; the Australian Research Council and the
Australian Department of Education, Science and Training;
the National Science Foundation of China and the Knowledge
Innovation Program of the Chinese Academy of Sciences under
contract No.~10575109 and IHEP-U-503; the Department of
Science and Technology of India; 
the BK21 program of the Ministry of Education of Korea, 
the CHEP SRC program and Basic Research program 
(grant No.~R01-2005-000-10089-0) of the Korea Science and
Engineering Foundation, and the Pure Basic Research Group 
program of the Korea Research Foundation; 
the Polish State Committee for Scientific Research; 
the Ministry of Education and Science of the Russian
Federation and the Russian Federal Agency for Atomic Energy;
the Slovenian Research Agency;  the Swiss
National Science Foundation; the National Science Council
and the Ministry of Education of Taiwan; and the U.S.\
Department of Energy.


\begin{thebibliography}{99}
\bibitem{chargeconjugate}
Charge-conjugate modes are included throughout this paper
unless noted otherwise.
\end{thebibliography}
\end{document}